\documentstyle[12pt,a4,psfig]{article}
\begin{document}
\baselineskip 0.8cm

\title{Contribution of Colour-singlet Process
$\Upsilon \to J/\Psi + c\bar{c}g$ to
$\Upsilon \to J/\Psi + X$}
\author{  Shi-yuan  Li$^a$,~
 Qu-bing  Xie$^b$,~    Qun Wang$^b$\\
\small{
$^a$ Department of Physics, Shandong University, Jinan, Shandong, 250100,
P. R. China}\footnote{e-mail: tpc@sdu.edu.cn}\\
\small{$^b$ Center of Theoretical Physics, CCAST(World Lab), 
Beijing,100080, P.R.
China $and$~ $^a$}}
\maketitle

\begin{abstract}
We show that the colour-singlet process $\Upsilon \to J/\Psi +
c\bar{c}g$  significantly contributes to the inclusive process
$\Upsilon \to J/\Psi + X$.
We calculate the partial width and the momentum 
distribution of the produced  
$J/\Psi$ in this channel.
The obtained width is comparable to  the experimental data.
The momentum distribution is  fairly soft,
which is in contrast to the results from the colour-octet processes
discussed earlier in literature but is consistent with 
the CLEO measurements.
Further experiments to check this contribution are suggested.
\end{abstract}

\newpage
Heavy quarkonium decay and production are  traditional
arena for QCD. The descriptions \cite{shu} of such processes 
are  nearly standard as 
a factorization procedure: 
The annihilation of $b\bar{b}$ or $c\bar{c}$ to gluon(s) 
or its creation from gluon(s)
can be calculated  via perturbative
QCD, since the  heavy quark mass  sets a hard scale for corresponding 
processes. The  bound state of the heavy quarkonium 
contains  non-perturbative effects.
This part is  `factorized out' and is described 
by QCD-inspired models \cite{ynd}.
Such a  procedure is  successful in describing many experimental data.
However, recently a large discrepancy for the prompt production 
of $J/\Psi$ and $\Psi'$ has been observed in Fermilab Tevatron between 
the data \cite{cdf} and the colour-singlet model prediction.
This lead to a re-inspection of the description of 
heavy quarkonium decay and production (for a review, see \cite{bfy}).
It was found that the `factorization' in colour-singlet model
had never been proved. And the colour-octet mechanism
 which is based on the 
NRQCD factorization scheme (see, e.g.\cite{nrqcd}) was proposed as a    
complement to the colour-singlet mechanism to explain
the above mentioned discrepancy.
It has been shown that 
the colour-octet mechanism plays an important r\^{o}le
in many  processes.
For example, it has been argued that two processes
$\Upsilon \to J/\Psi + gg$ \cite{cky} (see Fig.1(a)) and 
$\Upsilon \to J/\Psi + g$ \cite{nap} (see Fig.1(b)) where the 
colour-octet mechanism  plays the  key  r\^{o}le provide the leading
contribution to the inclusive process $\Upsilon \to J/\Psi + X$.
Experiments by the CLEO collaboration reported    
 a branching ratio of $(1.1\pm 0.4\pm 0.2) \times 10^{-3}$ and 
quite a soft $J/\Psi$ momentum spectrum (see dots in Fig.3)
 \cite{cleo}. 
The calculated branching ratios
of these two colour-octet processes  are indeed comparable 
to the data.
However,  the $J/\Psi$ momentum spectra of both processes
peak at  the largest value $\sim 4.2 GeV/c$,
which is in strong contradiction to the CLEO data.
(For $\Upsilon \to J/\Psi + gg$, see the dotted line in Fig.3; 
for $\Upsilon \to J/\Psi + g$, the $J/\Psi$ momentum spectrum is a 
$\delta$-function with the pole at $\sim 4.2GeV/c$.)

In this note,  we show that a colour-singlet process,   
$\Upsilon \to J/\Psi + c\bar{c}g$ (see Fig.2), also significantly 
contributes   to the inclusive
$J/\Psi$ production in $\Upsilon$ decay, and that 
the $J/\Psi$ momentum spectrum from this channel has to be soft,
which is consistent with the CLEO data. 
The latter point  can  be seen from the following simple kinematic
analysis.    
We recall that  angular momentum conservation and charge conjugation  
invariance  demand  that $\Upsilon$ decay mainly
through three (real or virtual) gluons.
It is clear that  $\Upsilon$ can
decay via one real  and two virtual gluons, and both of the virtual gluons
split into colour-octet $c\bar{c}$ (see Fig.2). 
In this case, a $c$ and a $\bar{c}$ originated from 
different gluons  can be in the colour-singlet, 
and they can form a $J/\Psi$ when they are in the proper 
angular momentum state and their invariant mass is near 
$J/\Psi$ mass shell.
Generally, the remaining $c\bar{c}$ created in 
this process  hadronizes
into open charm particles,
so the final particle system  excluding  $J/\Psi$ has a minimum  
invariant mass $2M_{D^0}$.
This implies that  the $J/\Psi$ momentum has a maximum of  3.3GeV/c
in this channel. 
From the Feynman 
diagram,
we see also no indication that the $J/\Psi$ momentum spectrum peaks
at this largest value, so it should be  soft.
The exact shape depends on the details of the dynamics which
is discussed later in this paper.
We also note that if only one instead of two  
 $c\bar{c}$  is created in the 3-gluon decay channel,
which  to the lowest orders in $\alpha_s$ correspond to the above mentioned 
two colour-octet processes \cite{cky, nap},  
the $J/\Psi$ momentum spectrum is very hard and in strong contrast to the
CLEO data, as  cited at the end of last paragraph.     
Hence we conclude  that in the  3-gluon decay channel and 
 to the lowest orders in $\alpha_s$,  
the process in Fig.2 is the only case
where the $J/\Psi$ momentum spectrum is soft.

The subsequent question  is then: how large is the 
branching ratio of this colour-singlet process?
At  first sight, one may expect that it is much smaller
than that of the colour-octet process
$\Upsilon \to J/\psi + gg$ discussed in \cite{cky},
since in the perturbative (short-distance) phase
the  colour-singlet process has
one more  $\alpha_s$ and one more virtual gluon propagator,
both of which give more  suppression.
But a careful inspection tells us that  it can be of
the same order of magnitude as that from the colour-octet
process $\Upsilon \to J/\psi + gg$, thus comparable to the data \cite{cleo}.   
The reason is as follows.
It is true that in the perturbative phase
the  colour-singlet process we consider receives more  suppression    
than the colour-octet process \cite{cky}. But  in the 
non-perturbative (long-distance) phase,
where the colour-singlet/octet $c\bar{c}$ form the $J/\psi$ resonance
with the  probability described by the 
colour-singlet/octet matrix elements \cite{nrqcd},  
the much larger colour-singlet matrix element  provides an enhancement.
Using the value of \cite{cky}, the colour-singlet matrix
element $<O^{J/\Psi}_{1}(^3S_1)>$ is
about 52 times as large as the colour-octet 
one  $<O^{J/\Psi}_{8}(^3S_1)>$.
If we use the more updated value in \cite{fhmn}, the ratio 
$\frac{<O^{J/\Psi}_{1}(^3S_1)>}{<O^{J/\Psi}_{8}(^3S_1)>}$ 
can even be 210 to 360.   
This enhancement can    
compensate  the perturbative suppression. So we  expect 
a non-negligible partial 
width of the  colour-singlet process $\Upsilon \to c\bar{c}(^3S_1, 1) 
+ c\bar{c}g \to J/\Psi + c\bar{c}g$,
where we use $(^3S_1, 1)$ to denote
 the quantum numbers of angular momentum
and colour for $c\bar{c}$.
Furthermore, if the invariant mass of 
the colour-singlet pair $c\bar{c}(^3S_1)$ is near the 
mass shell of $\Psi(2S)$, it can also form a $\Psi(2S)$ resonance, 
which decays into $J/\Psi$ with a branching ratio of $54.2\%$
\cite{pdg}. This indirect contribution $\Upsilon \to \Psi(2S)
+ c\bar{c}g \to J/\Psi + X$ 
provides another source of the inclusive 
$J/\Psi$ production in $\Upsilon$ decay.

Encouraged by these qualitative  analyses we calculate the partial
width of $\Upsilon$  and momentum distribution of $J/\Psi (\Psi(2S))$
for the direct process shown in Fig.2 and the process
$\Upsilon \to c\bar{c}(^3S_1, 1)
+ c\bar{c}g \to \Psi(2S) + c\bar{c}g$.   
Since both of them are  colour-singlet processes 
and the charm quark is heavy,
the treatment for the bound states 
can be a  conventional wave-function 
approach where the relative momentum between $c$ and $\bar{c}$
is vanishing, namely same as the case of positronium 
(see, e.g. \cite{nacht}). 
Similar to \cite{nacht},
the differential width of the direct process (see Fig.2) 
can be formulated as
\begin{equation}
\frac{d\Gamma}{dR}=\frac{|B_{J/\Psi} B_{\Upsilon} <c\bar{c}(^3S_{1},1)
c\bar{c}g| {\cal S} |b\bar{b}(^3S_{1},1)>|^2}{T}
\end{equation}
where $dR$ is the  8-dimensional phase space volume element for  
$J/\Psi$ and $c$, $\bar{c}, g$;
$\cal S$ is the S-Matrix; 
$B_{J/\Psi}$ and  $B_{\Upsilon}$ are related to 
the origin values of the wave functions 
of $J/\Psi$ and $\Upsilon$:
\begin{equation}
B_{J/\Psi}=\frac{\Psi^{*}_{J/\Psi}(0)}{\sqrt{m_c}},
\end{equation}
\begin{equation}
B_{\Upsilon}=\frac{\Psi_{\Upsilon}(0)}{\sqrt{V}2m_{b}}.
\end{equation}
For convenience, we normalize  all particle states to
$2EV$ (where $E$ is the particle's energy and $V$ is the volume of the total 
space) except for $\Upsilon$, which is normalized to  
1. This is why    
the coefficients of
the wave functions in (2) and (3) are different.
In (1) the sum over all spin states for  final particles  and average
of the 3 spin states for $\Upsilon$   are not explicitly shown and 
 the `time' $T$ is $2\pi\delta(0)$. The corresponding expression 
for the differential width of  the process  
$\Upsilon \to c\bar{c}(^3S_1,1)+ c\bar{c}g
\to \Psi(2S) + c\bar{c}g$ is the same. The 
only differences lie in  the phase space  and
the wave function at the origin, where  
the  corresponding quantities for 
$\Psi(2S)$ should be used to replace  those of  
$J/\Psi$.

For the perturbative sub-process $b\bar{b}(^3S_{1},1)
\to g^* g^* g \to c\bar{c} ~c\bar{c}~ g$, we consider,
as usual, only the six 
lowest-order Feynman diagrams at the tree-level
(see Fig.2). The treatment of 
the infrared behaviour of the real gluons in these diagrams 
is simple. It is much 
similar to that in the $\Upsilon \to 3g$ process (see, e.g. \cite{nrqcd}).
 For the  two diagrams where the real gluon vertex lies between
the two virtual
gluons, there is no infrared singularity due 
to `controlling momentum' \cite{nrqcd}.
For the rest four diagrams,  the infrared  gluon 
in each diagram leads to 
an eikonal vertex and an eikonal
 quark propagator \cite{nrqcd}.
These four diagrams form two pairs.
In each pair, the only difference between the two diagrams 
is the position of the real gluon vertex.  
The eikonal vertices in them have opposite signs, so 
 the infrared singularities of the two diagrams cancel
with each other.
Summing over all  the six diagrams, we obtain the 
infrared safe matrix element 
$<c\bar{c}(^3S_{1},1)c\bar{c}g| {\cal S} |b\bar{b}(^3S_{1},1)>$,
which is needed in calculating  $\frac{d\Gamma}{dR}$ from Equation(1).

To  get the numerical results, we need to know the value of 
$\alpha_s$ and those of the $\Psi_{J/\Psi}(0)$
and $\Psi_{\Upsilon}(0)$ (see Equation(2),(3)).
In order to compare the results with those 
of the colour-octet processes
\cite{cky,nap},
we use the same values for these parameters.
Especially, the scale of $\alpha_s$ is $M_{J/\Psi}$.  
The square of the radial wave function at
the origin for $\Psi(2S)$ have been 
calculated using different  potentials \cite{qig},  
Consistent with these calculations, we  use
$\frac{|{\cal R}_{\Psi(2S)}(0)|^2}{|{\cal R}_{\Psi(1S)}(0)|^2} \sim 1/2$
approximately, where ${\cal R}$ represents the radial wave function.
The quark masses are taken as
$m_{c}=\frac{M_{J/\Psi}}{2}$ and $m_{b}=\frac{M_{\Upsilon}}{2}.$
The partial width of the direct production process 
$\Upsilon \to c\bar{c}(^3S_{1},1) + c\bar{c}g \to
J/\Psi + c\bar{c}g$  is calculated to be 28.4eV.
Normalized by the full width of $\Upsilon$ \cite{pdg}, 
we obtain the corresponding  branching ratio of about
$0.54 \times 10^{-3}$. 
The indirect contribution to the branching ratio 
from $\Psi(2S)$ is small $\sim 0.05 \times 10^{-3}$.
So the branching ratio from these two sources 
is $0.59 \times 10^{-3}$,  
which is consistent with the CLEO result within the error.

If we take sum of the branching ratio obtained above for the 
colour-singlet process we consider,
and  those for the colour-octet processes discussed in 
\cite{cky,nap}, we obtain the total  branching 
ratio of $1.21 \times 10^{-3}$.
This result is still in agreement with CLEO data  (but is larger than
the ARGUS upper limit).
Because the value of the colour-singlet matrix elements can be extracted
from the pure leptonic decays of $\Upsilon$ and $J/\Psi$ etc.,
which are  electromagnetic process and 
where only the colour-singlet mechanism works,
the uncertainty is rather small.
This is in contrast to the 
case of extracting the value of 
the colour-octet matrix elements. The latter can  
only be extracted from $J/\Psi$ production  
experiments. It is therefore very sensitive to whether all the 
colour-singlet processes have been considered.
Furthermore, the perturbative part   
is sensitive to the  value of $\alpha_s$.
It is unclear whether $\alpha_s(M_{J/\Psi})$ 
or $\alpha_s(M_{\Upsilon})$ should be used \cite{nap}.
If $\alpha_s(M_{\Upsilon})$ instead of $\alpha_s(M_{J/\Psi})$
is used in both the colour-singlet process we considered here 
and the colour-octet processes considered in \cite{cky,nap}, 
the branching ratio of the colour-singlet process
 decreases to $0.12\times 10^{-3}$ which is still of the 
same order of magnitude as that of the colour-octet processes.
But the theoretical total width is 
smaller than the CLEO data.
It is now impossible to judge which mechanism
 is more important 
if we only look at their contributions to the branching ratio.

In contrast to the branching ratio,
the shape of the calculated 
$J/\Psi$ momentum spectrum does not depend on 
  the values of   
$\alpha_s$ and the non-perturbative
matrix elements. It is not influenced by 
the  uncertainties coming from these parameters.
Hence it can act as a `label' for a given decay channel. 
The $J/\Psi$ momentum distribution for the direct 
colour-singlet process(see Fig.2)
is shown in Fig.3 by the solid line. It has a peak near $2.2GeV/c$.  
 The shape is similar to the CLEO data. 
 About $90\%$ 
of the total events reside in the  range smaller than 3 GeV/c,
which is also consistent with the CLEO data.
The momentum spectrum of $\Psi(2S)$ is  shown by the dashed line,
 and the magnitudes of the differential width have been
multiplied by the branching ratio of $\Psi(2S) \to J/\Psi+X'$.
In the process $\Psi(2S) \to J/\Psi+X'$,
since $M_{\Psi(2S)}-M_{J/\Psi} << M_{J/\Psi}$, $J/\Psi$ carries most 
of the energy of $\Psi(2S)$. 
Hence the shape of the corresponding $J/\Psi$ momentum spectrum  is 
approximately the same as the $\Psi(2S)$.
We see from Fig.3 that the spectrum shape  
in the indirect process  is similar 
to that in the direct process.
For comparison, we also show in Fig.3
the momentum spectrum of the colour-octet process (see
Fig.1(a)) by the dotted line 
\footnote{We use Equation (24) in \cite{cky} to draw 
the line. We have changed the variable to $J/\Psi$ momentum
 and normalized the expression on the right hand side of the 
equation to the width for this channel \cite{cky}.}.
 We see clearly that 
the CLEO data favour the  colour-singlet process
considered in this note.

Finally, we would like to emphasize that it is not difficult
to check experimentally whether the colour-singlet process we consider
indeed plays  an important r\^{o}le for $\Upsilon \to J/\Psi+X$.
This is because the $J/\Psi$  in this colour-singlet  process  is 
definitely associated with  open charm particles  
and its  momentum is less than 3.3GeV/c.

In summary, both qualitative analyses and quantitative calculation
show that the colour-singlet process $\Upsilon \to J/\Psi + c\bar{c}g$
gives the branching ratio comparable to experimental data
and a similar soft $J/\Psi$ momentum spectrum as measured  
by CLEO Collaboration \cite{cleo} as well. 
Among the processes which have been studied theoretically 
up to now,
this is the only one which 
has such a feature. This seems to suggest 
that this process provide the  main contribution 
to the inclusive $J/\Psi$ production in $\Upsilon$ decay.

\vskip 60pt
 
ACKNOWLEDGEMENT

We thank Prof. Liang for carefully reading the manuscript 
and giving many helpful comments.
Li thanks Dr. Si (Aachen) for helpful discussions.
This work is supported in part by National Natural Science
Foundations of China (NSFC).

\newpage

Figure Captions

\vskip 30pt
Fig.1 Diagrams illustrating the colour-octet processes
$\Upsilon \to J/\Psi + gg$ \cite{cky} (a) and  
$\Upsilon \to J/\Psi + g$ \cite{nap} (b). 
\vskip 20pt
 
Fig.2 One of the six diagrams for the direct 
colour-singlet process $\Upsilon \to
c\bar{c}(^3S_1, 1) + c\bar{c}g \to J/\Psi + c\bar{c}g$. 
The real gluon vertex can be between  those of the two virtual gluons or
in either side of them. And the two virtual gluons can change their order.
\vskip 20pt

Fig.3 The momentum spectrum of $J/\Psi$ in the direct 
colour-singlet process we consider  (solid line) 
and the momentum spectrum of $\Psi(2S)$ 
multiplied by the branching ratio of
$\Psi(2S) \to J/\Psi + X'$  for the 
process $\Upsilon \to \Psi(2S) +c\bar{c}g$
(dashed line).  The CLEO data \cite{cleo} is shown by dots.  
The result of the colour-octet process of
 Fig.1(a) is shown for comparison (dotted line).

\newpage

\vskip 30pt
\begin{figure}
\psfig{file=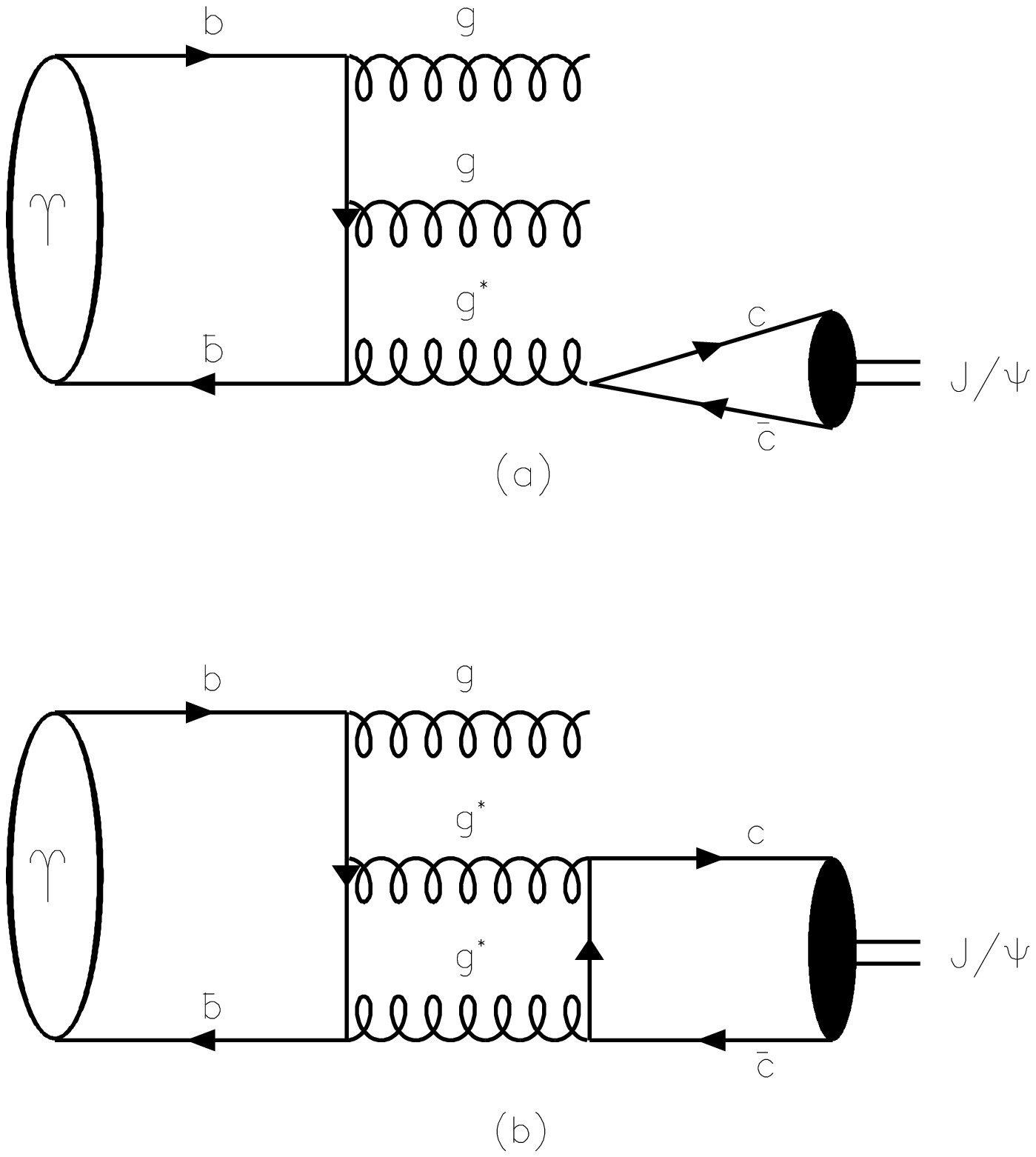,height=10cm,width=10cm}
\caption{~}
\end{figure}

\vskip 30pt
\begin{figure}
\psfig{file=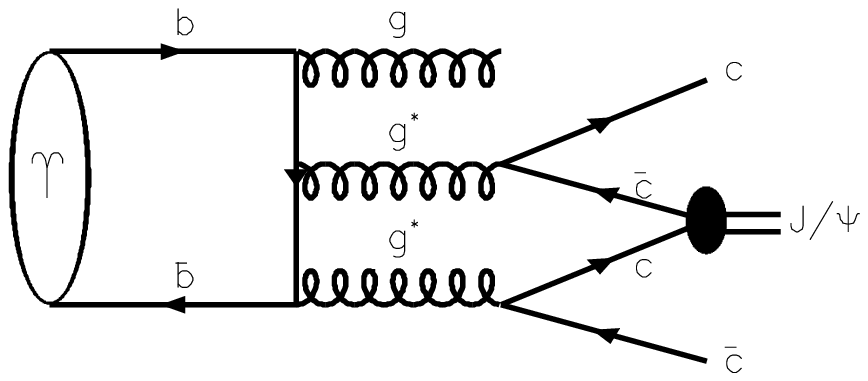,height=8cm,width=14cm}
\caption{~}
\end{figure}
 
\vskip 30pt
\begin{figure}
\psfig{file=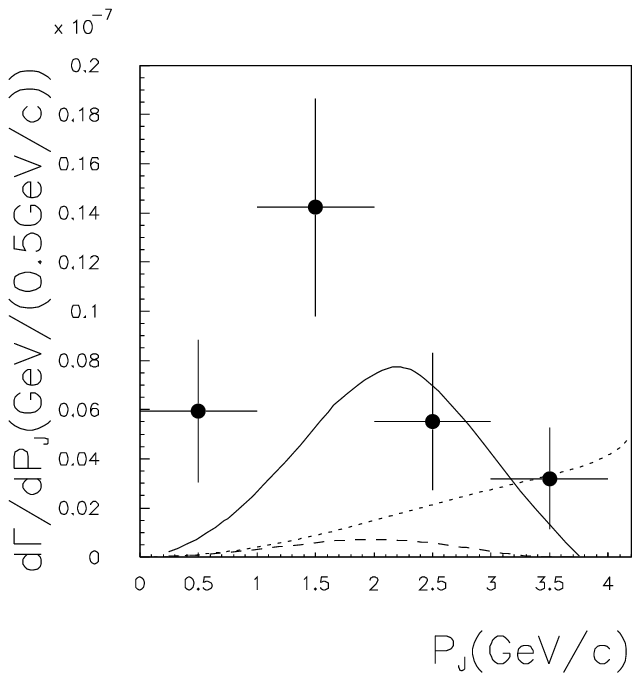,height=9cm,width=9cm}
\caption{~}
\end{figure}

\end{document}